\documentclass{iopart}
\usepackage{cite}
\jl{6}        
\eqnobysec    

\def\beq{\begin{equation}}
\def\eeq{\end{equation}}

\def\dualp#1{{}^{\ast_{(\hbox{$\scriptstyle #1$})}} \kern-1pt}

\catcode`@=11
\def\meqalign#1{\null\,\vcenter{\openup\jot\m@th
  \ialign{\strut\hfil$\displaystyle{##}$&&$\displaystyle{{}##}$\hfil
      \crcr#1\crcr}}\,}
\catcode`@=12

    \let\Box=\dal
\def\rmd{{\rm d}}

\def\pmb#1{\setbox0=\hbox{$#1$}%
  \kern-.025em\copy0\kern-\wd0
  \kern.05em\copy0\kern-\wd0
  \kern-.025em\raise.0433em\box0}
\def\bfrho{{\pmb{\rho}}} \def\bfgamma{{\pmb{\gamma}}}
\def\bfnabla{{\pmb{\nabla}}}

\input prepictex \input pictex \input postpictex
\def\mathput#1{\relax \ifmmode \displaystyle #1\else $\displaystyle #1$\fi}

\begin{document}

\title[Gravitomagnetism in the
KNTN spacetime]{Gravitomagnetism in the Kerr-Newman-Taub-NUT spacetime}

\author{Donato Bini\dag ${}^\ast$, Christian Cherubini\ddag ${}^\ast {}^\diamond$,\\
Robert T. Jantzen\S ${}^\ast$, Bahram Mashhoon\P}

\address{\dag\  Istituto per le Applicazioni del Calcolo \lq\lq M. Picone", C.N.R.,
 I-- 00161 Roma, Italy}
\address{${}^\ast$ ICRA International Center for Relativistic Astrophysics,
University of Rome, I--00185
Rome, Italy}
\address{\S\  Department of Mathematical Sciences, Villanova University,
  Villanova, PA 19085, USA}
\address{\ddag\ Dipartimento di Fisica ``E.R.
Caianiello'', Universit\`a di Salerno, I--84081, Italy}
\address{\P\ Department of Physics and Astronomy,
University of Missouri-Columbia, Columbia,
Missouri 65211, USA}
\address{${}^\diamond$ Institute of Cosmology and Gravitation, University of Portsmouth, Portsmouth PO1 2EG, England, UK}

\begin{abstract}
We study the motion of test particles and electromagnetic waves in the
Kerr-Newman-Taub-NUT
spacetime in order to elucidate some of the effects associated with the
gravitomagnetic
monopole moment of the source. In particular, we determine in the linear
approximation the
contribution of this monopole to the gravitational time delay and the
rotation of the plane
of the polarization of electromagnetic waves.
Moreover, we consider \lq\lq spherical" orbits of uncharged test 
particles in the Kerr-Taub-NUT spacetime and discuss 
the modification of the Wilkins orbits 
due to the presence of the gravitomagnetic 
monopole.
\end{abstract}


\submitted 28 August 2002 [appeared Vol.~2, 457--468 (2003)]

\section{Introduction}

In its simplest manifestation, gravitomagnetism is caused by the current of
mass-energy in direct analogy
with electrodynamics. In particular, gravitomagnetic effects can be found
in the external gravitational
field of rotating astronomical masses.

In this article we are interested in the exterior Kerr-Newman-Taub-NUT (KNTN)
solution given in the Boyer-Lindquist coordinates by the metric
\cite{DemNew,Mil,NUT,McGRuf,DodTur}
\begin{eqnarray}
\label{metrica}
\rmd s^2
   &=&-\frac{1}{\Sigma}(\Delta -a^2\sin^2\theta)\rmd t^2
    +\frac{2}{\Sigma}[\Delta \chi -a(\Sigma +a \chi)\sin^2\theta]\rmd t
\rmd \phi \nonumber \\ 
    &&+\frac{1}{\Sigma}[(\Sigma +a \chi)^2\sin^2\theta -\chi^2 \Delta ]\rmd \phi^2
    +\frac{\Sigma}{\Delta}\rmd r^2 +\Sigma d\theta^2
\end{eqnarray}
and the corresponding electromagnetic Faraday tensor can be expressed in terms of the 2-form
\begin{eqnarray}
\label{max}
F&=&\frac{Q}{\Sigma^2}\{ [r^2-(\ell+a\cos\theta)^2]\rmd r \wedge (\rmd t
-\chi\rmd \phi)\nonumber \\
    && +2r\sin\theta (\ell+a\cos\theta)\rmd \theta \wedge
[(r^2+a^2+\ell^2)\rmd \phi -a \rmd t]\} .
\end{eqnarray}
Here  $\Sigma$, $\Delta$, and $\chi$ are defined by
\beq
\fl
\Sigma = r^2 +(\ell+a\cos\theta)^2,\ 
\Delta = r^2-2Mr-\ell^2 +a^2+Q^2,\ 
\chi = a \sin^2\theta -2\ell\cos\theta\ .
\eeq
Units are chosen such that $G=c=1$, so that $(M,Q,a,\ell)$ all have the
dimension of length:
the source has mass $M$, electric charge $Q$, angular momentum $J=Ma$
(i.e., gravitomagnetic dipole
moment) along the $z$-direction, and gravitomagnetic monopole moment
$\mu=-\ell$,
where $\ell$ is the NUT parameter. In this family the charged KNTN metric can be
obtained from the uncharged
Kerr-Taub-NUT (KTN) metric by the simple transformation $M\to M-Q^2/(2r)$
in $\Delta$, while the Taub-NUT
metric corresponds to $a=0$
\cite{Mis,MisTau,Bon,Fei}.
It follows from equation (\ref{max}) that in the far zone ($r\gg a$ and $r\gg |\ell|$), the lowest-order contribution to the electric field is the radial Coulomb field, while the lowest-order contribution to the magnetic field is due to the field generated by the magnetic dipole moment $Qa$ of the source as well as terms of similar order proportional to $2Q\ell$.

An interesting example of a gravitomagnetic phenomenon is provided by the
clock effect for circular
equatorial orbits around rotating masses \cite{bjm,bjm2}. To illustrate the
situation in a more general
context, consider the circular equatorial motion of a test particle of mass
$m$ and negative electric
charge $-q<0$ around the Kerr-Newman black hole of mass $M$, charge
$Q>0$, and angular momentum
$J=Ma$. Here we neglect all (electromagnetic and gravitational) radiative
effects for the sake of
simplicity. A Newtonian analysis implies that the test particle
moves along a circle of radius
$r$ with uniform angular frequency given by $(\frac{M+\eta Q}{r^3})^{1/2}$,
where $-\eta=-q/m<0$ is the
charge-to-mass ratio of the test mass. This Newtonian result is expected to
emerge from the following
general relativistic analysis when the test particle is very far from the black hole. 
In the relativistic theory, the Lorentz equation of motion
\beq
\label{motion}
\frac{\rmd^2 x^\mu}{\rmd \tau^2}
+\Gamma^\mu{}_{\alpha\beta} \frac{\rmd x^\alpha}{\rmd \tau} \frac{\rmd
x^\beta}{\rmd \tau}
=-\eta F^\mu{}_\nu \frac{\rmd  x^\nu}{\rmd \tau}
\eeq
must be solved. Here $\tau$ is the proper time of the test particle along
its world line. A detailed
analysis reveals that in this case Eq.~(\ref{motion}) reduces to
\begin{eqnarray}
\label{sys1}
\fl\qquad
&&\omega_K^2 \left(\frac{\rmd t}{\rmd \tau} -a \frac{\rmd \phi}{\rmd \tau}\right)^2 
-\left(\frac{\rmd \phi}{\rmd \tau}\right)^2
=\omega_C^2 \left(\frac{\rmd t}{\rmd \tau} -a \frac{\rmd \phi}{\rmd \tau}\right)
\ ,\\ \label{sys2}
\fl\qquad
&&\left(\frac{\rmd t}{\rmd \tau}\right)^2
- \left(\frac{2M}{r}-\frac{Q^2}{r^2}\right)
\left(\frac{\rmd t}{\rmd \tau} -a \frac{\rmd \phi}{\rmd \tau}\right)^2
-(r^2+a^2)\left(\frac{\rmd \phi}{\rmd \tau}\right)^2 =1
\ ,
\end{eqnarray}
where $\pm\omega_K$ ($\omega_K>0$) defined by
\beq
\omega_K^2 =\frac{M}{r^3}-\frac{Q^2}{r^4}
\eeq
is the Keplerian frequency for a neutral test particle in circular orbit
around a
Reissner-Nordstr\"om black hole and
$\pm \omega_C$ ($\omega_C\ge0$) defined by
\beq
\omega_C^2=\eta \frac{Q}{r^3}
\eeq
would be the frequency of motion of the test particle of charge $-\eta m$
if only the Coulomb interaction
were present.

To analyze the system (\ref{sys1})--(\ref{sys2}), first set $a=0$; then 
$(d\phi /dt)^2$ satisfies a
quadratic equation that has two possible solutions. These can be
distinguished by the fact that in one
case $dt/d\tau >0$ and in the other case $dt/d\tau <0$. We must impose the
physical requirement that the
time coordinate $t$ increases along timelike world lines. With this
postulate, one then obtains
$\rmd \phi/\rmd t=\pm \Omega$,
where $\Omega>0$ is given by
\beq
\Omega^2=\omega_K^2 -\frac12 r^2\omega_C^4+N\omega_C^2
\eeq
with
\beq
N=\sqrt{1-\frac{3M}{r}+\frac{2Q^2}{r^2}(1+\frac{\eta^2}{8})} \ .
\eeq
Here $\Omega^2$---in contrast to the second solution that we have ruled
out---properly reduces to the
Newtonian result far from the black hole.

Next consider a charged rotating black hole ($a\neq0, Q\neq0$).
The results can be simply stated  for $a/M \ll 1$, in which case
to first order in $a/M$ one finds
\beq
\frac{\rmd t}{\rmd \phi} =\pm \frac{1}{\Omega}+ a (1-\mathcal{Q}) \ ,
\eeq
where
\beq
\mathcal{Q}
=\frac12 \frac{\eta Q}{N[M-\frac{Q^2}{r}(1+\frac12 \eta^2)+ \eta Q N]}\ .
\eeq
Thus if $t_+$ ($t_-$) denotes the coordinate time period for a complete
revolution around the black hole in the counterclockwise (clockwise) sense
($\Delta\phi=\pm2\pi$), then
\beq
t_+ -t_-=4\pi a (1-\mathcal{Q}) \ ,
\eeq
which is the expression for the single-clock clock effect \cite{bjm,bjm2} in
this case. Note that as $r \to
\infty$
this result reduces to
\beq
t_+ -t_-=4\pi a \frac{M+\frac12 \eta Q}{M+\eta Q} \ ,
\eeq
so that the effect is always less than $4\pi a$, the result for a
neutral particle.
It is interesting to investigate the two-clock clock effect \cite{bjm,bjm2} in
this case as well, i.e.
$\tau_+ -\tau_-$, where $\tau$ is the corresponding proper time as measured
along the geodesics. One finds
\beq
\frac{\rmd \tau}{\rmd \phi}
=\pm \frac{\Omega^2-\omega_K^2}{\Omega \omega_C^2}+a (1-\mathcal{Q'}) \ ,
\eeq
where
\beq
1-\mathcal{Q'}=\frac12 N^{-1}(1+\frac{\omega_K^2}{\Omega^2}) \ .
\eeq
It follows that
\beq
\tau_+-\tau_- =4\pi a (1-\mathcal{Q'})
\eeq
and that $\mathcal{Q}'\to \mathcal{Q}$ as $r\to \infty$.
These results, valid only to first order in $a/M$, indicate a certain
gravitomagnetic temporal structure around
the Kerr-Newman black hole that is consistent with the axial symmetry of
the underlying spacetime
structure. It is the purpose of this paper to clarify how gravitomagnetism
is affected by the presence of
the gravitomagnetic {\it monopole} moment $\mu =-\ell $, which has
{\it spherical} symmetry.

In the following section, the gravitomagnetic field is studied and the
results are applied in section 3 to
the propagation of test electromagnetic waves in the Kerr-Taub-NUT
spacetime. Section 4 describes
circular holonomy in this spacetime. Spherical orbits are considered in
section 5. This is followed in the
final section by a discussion of our results.

\section{Gravitomagnetic field}

Even though the spacetime curvature vanishes asymptotically at large
spatial distances,
the KNTN spacetime is not asymptotically flat when
$\ell\neq0$,
a fact
intimately related to the gravitomagnetic monopole character of the source
associated with this
parameter. At present there is no evidence for the existence of either
magnetic or gravitomagnetic
monopoles. Moreover, the lack of asymptotic flatness for a spacetime
describing a localized source
appears to be unphysical.
However,
like the G\"odel spacetime, the Kerr-Newman-Taub-NUT spacetime can play a
useful role in clarifying
important gravitomagnetic features of classical general relativity.

With this goal in mind, we set $Q=0$
and
expand the metric to first order in $a$ and $\mu=-\ell$; in this way one
can more easily interpret both
parameters in the context of gravitomagnetism. One finds
\begin{eqnarray}\label{eq:met}
\rmd s^2
 &=& -\left(1-\frac{2M}{r}\right) \rmd t^2
 + \left(1-\frac{2M}{r}\right)^{-1} \rmd r^2 + r^2 (
\rmd\theta^2+\sin^2\theta\, \rmd\phi^2)
\nonumber\\
&& -4\left[ \frac{aM}{r} \sin^2\theta - \mu \cos\theta
\left(1-\frac{2M}{r}\right) \right]
                        \rmd t \rmd \phi
\ ,
\end{eqnarray}
namely, the Schwarzschild metric plus the Lense-Thirring and Taub-NUT terms.
While the time-coordinate  slices are intrinsically asymptotically flat, the
fact that
$g_{t\phi}\to -2\ell\cos\theta$ as $r\to\infty$ implies that the 
spacetime is not asymptotically flat~\cite{Mis,MisTau}.

By introducing the isotropic radial coordinate $\rho$ defined by
$r=\rho(1+\Phi/2)^2$, where $\Phi=M/\rho$ is the (sign-reversed) Newtonian
gravitational potential,
and the corresponding isotropic Cartesian-like coordinates
$x=\rho\sin\theta\cos\phi$, $y=\rho\sin\theta\sin\phi$, and $z=\rho\cos\theta$ 
(in 3-vector form $\bfrho  = \rho
\hat{\bfrho} =(x,y,z)$),
the metric takes the form
\begin{eqnarray}\label{eq:TNquasi}
\rmd s^2
 &=& -\left(\frac{1-\Phi/2}{1+\Phi/2}\right)^2 \rmd t^2
 + (1+\Phi/2)^4\,  (\rmd x^2 + \rmd y^2 +\rmd z^2)
\nonumber\\
&& -4\mathbf{A}_{\rm(g)}\cdot \rmd \bfrho\, 
                        \rmd t
\ .
\end{eqnarray}
We interpret this metric as representing a background Minkowski spacetime
in inertial coordinates
$(t,x,y,z)$ together with a source characterized by a gravitoelectric
potential $\Phi$ and a linearized
gravitomagnetic vector potential $\mathbf{A}_{\rm(g)}$ given by
\beq
\mathbf{A}_{\rm(g)}= \mathcal{F}(\mathbf{\hat J} \times \bfrho )
\leftrightarrow
\mathbf{A}_{\rm(g)}\cdot \rmd \bfrho
= \mathcal{F}\rho^2\sin^2\theta\; \rmd\phi
\eeq
(i.e., the slicing gravitomagnetic potential 1-form $N_a \rmd x^a=g_{ta}
\rmd x^a$ when multiplied by
$-2$, see \cite{mfg}). 
Here
$\mathbf{J} = Ma\, \mathbf{\hat z}$ is the angular momentum, 
\beq
\mathcal{F} = \frac{aM}{\rho^3(1+\Phi/2)^2}
 - \frac{\mu z}{\rho(x^2+y^2)} \left(\frac{1-\Phi/2}{1+\Phi/2}\right)^2
\ ,
\eeq
and the 3-vector operations are carried out in the flat spatial metric,
so that
\beq
(\mathbf{\hat z} \times \bfrho)\cdot \rmd \bfrho =\rho^2\sin^2\theta\; \rmd
\phi =(\rho^2-z^2)\rmd \phi\ .
\eeq

In the case that $M/\rho \ll 1$ far from the source, one has
\beq
  \mathcal{F}\approx \frac{Ma}{\rho^3} - \frac{\mu z}{\rho(x^2+y^2)}\ ,\
  \mathbf{A}_{\rm(g)} \approx \frac{\mathbf{J}\times \bfrho }{\rho^3}
                - \frac{\mu z\,\mathbf{\hat z} \times \bfrho }{\rho(x^2+y^2)}\ ,
\eeq
and the gravitomagnetic vector field $\mathbf{B}_{\rm(g)}
=\pmb{\nabla}\times \mathbf{A}_{\rm(g)}$ is given by
\beq\label{eq:Bg}
 \mathbf{B}_{\rm(g)}
   = \frac{J}{\rho^3}
[3(\hat{\bfrho}  \cdot \mathbf{\hat J}) {\hat \bfrho} - \mathbf{\hat J}]
          + \frac{\mu \bfrho }{\rho^3} \ .
\eeq
Thus $J$ corresponds to the gravitomagnetic dipole moment and $\mu$ to the
gravitomagnetic monopole
moment. The threading gravitomagnetic vector potential (i.e., the 1-form
$M_a \rmd x^a=-(g_{ta} /g_{tt}) \rmd x^a$, see \cite{mfg}) defines another vector
\beq
\label{bfgamma}
\fl\qquad
   \bfgamma = -2 \mathcal{F}\left(\frac{1+\Phi/2}{1-\Phi/2}\right)^2
\, \mathbf{\hat J} \times \bfrho
            = - 2\frac{\mathbf{J}\times\bfrho }{\rho(\rho-M/2)^2}
         +2\frac{\mu z\,\mathbf{\hat z} \times\bfrho }{\rho(x^2+y^2)}\ ,
\eeq
so that $\bfgamma \approx -2 \mathbf{A}_{\rm (g)}$ for $\rho \gg M$.

One can now use the metric (\ref{eq:TNquasi}) to treat phenomena in terms
of a gravitoelectric field
$\mathbf{E}_{\rm(g)}=-\bfnabla \Phi$ and a gravitomagnetic field
$\mathbf{B}_{\rm(g)}=-\bfnabla \Psi$, where
\beq
\Psi = \frac{\mathbf{J}\cdot \bfrho}{\rho^3}+\frac{\mu }{\rho}
\eeq
is the gravitomagnetic scalar potential \cite{mas93}. In the definition of
gravitoelectric and
gravitomagnetic fields, certain conventional numerical factors of $2$, etc., are
unavoidable depending on the
specifics of the analogy with electromagnetism.
It turns out that in the description of the linearized gravitational motion
of a test particle of inertial
mass $m$ around a central source of inertial mass $M$, one can employ the
{\it same} formulas as in
classical electrodynamics if one assigns a gravitoelectric charge $M$ and
a gravitomagnetic charge $2M$
to the central source and corresponding gravitational charges of $-m$
and $-2m$ to the test mass \cite{mas93}.  The different signs ensure
that gravitation is attractive; moreover, the ratio of gravitomagnetic
charge to the gravitoelectric charge is always $2$ due to the spin-$2$ 
character of the linearized gravitational field.  We employ this
convention in the present work.

Geodesic motion of test particles in the field of a gravitomagnetic
monopole has been considered in
\cite{ZimSha,LynNou} and in references cited there, while lensing
aspects of a gravitomagnetic monopole have been discussed briefly in
\cite{LynNou}.  In the following section we consider the effect of this
monopole on the propagation of electromagnetic waves.

\section{Wave propagation}

Consider the propagation of electromagnetic waves in the Kerr-Taub-NUT
spacetime. We work to linear order
in $a$ and $\mu$ and express the electromagnetic perturbations in terms of
the background Cartesian-like
coordinates $(t,x,y,z)$ introduced above. Writing the sourcefree Maxwell
equations in these coordinates
as $F_{[\alpha\beta,\gamma]}=0$ and $[\sqrt{-g}
F^{\alpha\beta}]_{,\beta}=0$ and introducing the standard
decompositions $F_{\alpha\beta}\to (\mathbf{E},\mathbf{B})$ and
$\sqrt{-g} F^{\alpha\beta} \to (-\mathbf{D},\mathbf{H})$ \cite{Skr}, one
finds the equivalent flat
spacetime Maxwell equations in inertial coordinates in the presence of
an effective material medium \cite{Skr,Ple}, i.e., 
\beq 
\mathbf{\nabla} \cdot \mathbf{D} = 0\ ,\ 
\mathbf{\nabla} \cdot \mathbf{B} = 0\ ,\
\mathbf{\nabla} \times \mathbf{E} = -\partial_t \mathbf{B}\ ,\
\mathbf{\nabla} \times \mathbf{H} = \partial_t \mathbf{D}\ .  
\eeq
Therefore we may think of the electromagnetic field as propagating in
an inertial frame in Minkowski spacetime but in the presence of a
gravitational ``medium" with certain constituitive properties given by
\cite{Skr,Ple,cohmas,mas75} 
\beq 
D_i = \epsilon_{ij} E_j -
[\bfgamma\times\mathbf{H}]_i\ ,\ B_i = \mu_{ij} H_j +
[\bfgamma\times\mathbf{E}]_i\ , 
\eeq 
where $ \epsilon_{ij} = \mu_{ij}
= -\sqrt{-g} g^{ij}/g_{tt}$ and $\gamma_i = - g_{ti}/g_{tt}$ 
( $=M_i$,
the threading shift 1-form as above).  For the present metric one has
\beq 
\epsilon_{ij} = \mu_{ij} = \mathcal{N} \delta_{ij}\ ,\
\mathcal{N} = \frac{(1+\Phi/2)^3}{1-\Phi/2}\ , 
\eeq
and $\bfgamma$ is given by Eq.~(\ref{bfgamma}).
Here $\mathcal{N}$ can be interpreted as an effective index of refraction
of the gravitational medium and
$\bfgamma$
represents the gyrotropic aspects of the medium. We note that $\bfgamma$
is singular along the entire $z$-axis due to the presence of the term 
involving $\mu \not =0$.
However, this singularity can be removed by a local coordinate 
transformation in any given exterior neighborhood \cite{DemNew,Mis}, 
a fact that is consistent with the gauge dependence of the vector potential.

Since these constitutive quantities are all independent of time, the
electromagnetic waves propagate with
constant frequency $\omega$ in this gravitational medium.  With a
time dependence of the form $\exp (-i\omega t)$, Maxwell's equations then
reduce
to the wave equation \beq\label{eq2.13} (i^{-1}\bfnabla - \omega
\bfgamma)\times
\mathbf{W}=-i\omega \mathcal{N} \mathbf{W} \eeq for the complex
Kramers vector $\mathbf{W}=\mathbf{E}+ i \mathbf{H}$.
Note that the gravitomagnetic contribution to this Dirac-type equation
enters via the substitution
$i^{-1}\bfnabla  \to i^{-1}\bfnabla - \omega \bfgamma$. Therefore, it
follows from the Aharonov-Bohm
effect that radiation following a closed spatial path $C$---via reflection
from devices such as
mirrors, for example (see Fig.~1)---in the exterior field of a gravitomagnetic source
exhibits the following phase difference comparing the two waves after
one loop in the counterclockwise ($+$) direction and one loop in the
clockwise ($-$) direction
along the closed path $C$ \cite{cohmas}
\beq\label{eq:phase}
  \varphi_+ - \varphi_- = 2 \omega \oint_{C} \bfgamma \cdot \rmd \bfrho\ .
\eeq
For $\rho \gg M$, $\bfnabla\times\bfgamma\simeq -2{\bf B}_{\rm(g)}$; therefore,
one finds that $\varphi_+ - \varphi_-$ is $-4\omega$ times the
gravitomagnetic flux threaded by the closed path $C$.

\begin{figure}
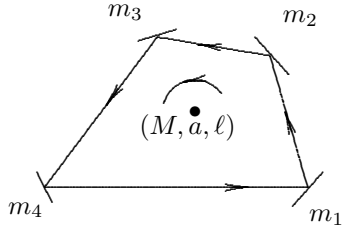

\typeout{Figure 1}
$$ \vbox{\beginpicture
\setcoordinatesystem units <.5cm,.5cm> point at  0 0
\put {\mathput{\bullet}}                 at  -1 0
\put {\mathput{(M,a,\ell )}}                   at  -1.2 -0.4
\put {\mathput{m_1}}       at 2.5 -2.8
\put {\mathput{m_2}}       at 1.8 2.5
\put {\mathput{m_3}}       at -2.85 2.65
\put {\mathput{m_4}}       at -5.5 -2.6
\plot -5 -2 -2 2 /
\plot -2 2 1 1.5 /
\plot 1 1.5 2 -2 /
\plot -5 -2 2 -2 /
\plot -4.8 -2.4 -5.2 -1.6 /
\plot -2.5 1.9   -1.5 2.2 /
\plot 0.6 2 1.5 1 /
\plot 1.6 -2.5  2.4 -1.6 /
\circulararc 130 degrees from -.3 .5 center at -1 0

\arrow <.3cm> [.1,.3] from  1.57 -0.5 to 1.4 0
\arrow <.3cm> [.1,.3] from  -0.5 1.75 to -0.9 1.8
\arrow <.3cm> [.1,.3] from  -3.05 0.6 to -3.35 0.2
\arrow <.3cm> [.1,.3] from  0 -2 to  0.4 -2

\arrow <.3cm> [.1,.3] from  -1 0.85 to  -1.3 0.8
\endpicture}$$
\caption{Propagation of null rays along a closed contour around a mass
$M$ with gravitomagnetic monopole $(\mu =-\ell)$ and dipole $(J=Ma)$
moments.  The ``mirrors'' $m_1$--$m_4$ in this schematic diagram could be
transponders on-board artificial satellites.}
\label{fig:1}
\end{figure}

If we consider matter waves propagating around this path, corresponding to
radiation of particles
with nonzero rest mass $m>0$ instead of photons, then $\omega$ in
Eq.~(\ref{eq:phase}) must be
interpreted as the de Broglie frequency of the matter waves, i.e.
$\omega\to m/\hbar$.
Moreover, a different treatment is required if instead of $C$ a path defined
by optical fibers (or a ring
laser) is employed.  Clearly the constitutive properties of the
corresponding medium should then be taken into account as well.

If the dominant wavelength of the waves is negligible compared to the
characteristic length scales of the
medium, one can interpret Eq.~(\ref{eq:phase}) in terms of rays of radiation.
In this case, let $P$ be an
observer at a point on the closed path $C$. Then according to this observer
the interval of the coordinate
time  $t_+$ ($t_-$) for the rays to go around $C$ in the positive
(negative) sense is such that
$\varphi_+ - \varphi_- = \omega(t_+-t_-)$. Thus
\beq
  t_+-t_- = 2 \oint_{C} \bfgamma \cdot \rmd \bfrho\ ,
\label{tpiutmeno}
\eeq
which is reminiscent of the Sagnac effect via the gravitational Larmor theorem
\cite{mas93}.

Consider now a closed path $C$ in the $(x,y)$-plane far from the source
($\rho \gg M$) and note that the
monopole contribution to $t_+-t_-$ in Eq.~(\ref{tpiutmeno}) vanishes due to
the fact that $z=0$ in this plane (see Eq.~(\ref{bfgamma})). 
It should also vanish in any other plane through the origin by the
spherical symmetry of the
monopole. This is clear from a simple application of Stokes' theorem
(see below) if $C$ does not
encircle the singularity at $\rho =0$. If it does, then let $C_0$ be a
circle in this plane around the
origin, such that $C$ is always outside $C_0$. It follows from Stokes'
theorem that
\beq
\oint_{C-C_0} \frac{z [\mathbf{\hat z} \times \hat \bfrho ] \cdot \rmd
\bfrho}{x^2+y^2}= - \int \frac{\bfrho \cdot \rmd \mathbf{S}}{\rho^3}=0
\eeq 
since $\bfrho $ in this plane is orthogonal to the area element
$\rmd \mathbf{S}$ by definition.  It now remains to calculate 
\beq
I=\oint_{C_0} \frac{z [\mathbf{\hat z} \times \hat \bfrho ] \cdot
\rmd \bfrho}{x^2+y^2} 
\eeq 
in this plane and to show that it vanishes
regardless of the radius $r_{0}$ of the circle $C_0$.  Taking
advantage of the axial symmetry of the ``medium'', it suffices to
calculate $I$ in the $(x',y')$-plane, where the $(x',y',z')$ coordinate system is obtained
from the $(x,y,z)$ system by a rotation of angle $\alpha$ about the
$x$-axis.  Thus 
\beq x=x',\quad y=y'\cos\alpha - z' \sin \alpha, \quad
z=y'\sin \alpha + z' \cos\alpha .  
\eeq 
A straightforward calculation
then results in 
\beq 
I=\sin\alpha \cos \alpha \int_0^{2\pi} \frac{\sin
\phi\; \rmd \phi}{\cos^2\alpha +\sin^2\alpha \cos^2\phi}=0 \ ,
\eeq 
where
$x'=r_{0}\cos\phi$, $y'=r_{0}\sin \phi$, and $z'=0$ along $C_0$.  

It is
interesting to point out that an equation very similar to Eq.~(\ref{eq2.13})
would hold in a background with Cartesian-like coordinates for the
propagation of electromagnetic waves in the general Kerr-Taub-NUT
spacetime.  In this spacetime, one must identify $t$ with
$t+n(8\pi\ell)$ for any integer $n$; that is, $t$ is assumed to be
periodic with period $8\pi\ell$ \cite{DemNew,Mis}.  Thus 
in the presence of a gravitomagnetic monopole moment a temporal delay equation of the
form (3.6) should be interpreted modulo an integer multiple of $8 \pi \ell$; moreover,
a time
dependence of the form $\exp(-i\omega t)$ implies that $4\omega \ell$
must be an integer.  Hence the photon energy is quantized in units of
$ \hbar /(4\ell)$.  Various aspects of this quantization condition have
been explored in \cite{Dow,MuePe}.  We are interested in very high
frequency radiation; therefore, this discreteness will not be pursued
further here.

Let us note that in the JWKB limit, one can solve the wave
equation (\ref{eq2.13}) with the ansatz
\beq 
\mathbf{W}=\mathbf{W}_0\, e^{-i\omega (t-t_0)+i\omega
S(\mathbf{x})},\label{eq2.20}
\eeq
where $\mathbf{W}_0=\mathbf{W}(t_{0},\mathbf{x}_{0})$ and
\beq 
S(\mathbf{x})
=\int^{\mathbf{x}}_{\mathbf{x}_0}\bfgamma
(\mathbf{x}')\cdot \rmd\mathbf{x}'
+\int^{\mathbf{x}}_{\mathbf{x}_0}\mathcal{N}({\bf x}')\, \hat \mathbf{k}
\cdot
\rmd \mathbf{x}'.
\eeq
Here $\hat{\mathbf{k}}$ is a constant unit vector 
that indicates the direction of wave propagation in the  background inertial frame.
It is useful
to write the eikonal $S$ in the form
\beq 
\label{eq2.22} 
S(\mathbf{x})=\hat{\mathbf{k}}\cdot
(\mathbf{x}-\mathbf{x}_0)+\int^{\mathbf{x}}_{\mathbf{x}_0}[\mathcal{N}(\mathbf{x}')-1]\, 
\hat {\mathbf{k}}\cdot
\rmd \mathbf{x}'+\int^{\mathbf{x}}_{\mathbf{x}_0} \bfgamma (\mathbf{x}')\cdot
\rmd \mathbf{x}',
\eeq
where the gravitational delay effects are separated from the simple
plane-wave propagation aspects of
the solution. Moreover, let $\hat{\mathbf{k}}$, $\hat{\mathbf{n}}_1$, and
$\hat{\mathbf{n}}_2$ form an
orthonormal triad in the background Euclidean space such that
$\hat{\mathbf{k}}\times
\hat{\mathbf{n}}_1=\hat{\mathbf{n}}_2$. Then
$\mathbf{W}_0=A( \hat{\mathbf{n}}_1+i\hat{\mathbf{n}}_2 )$,
where $A$ is a complex amplitude.

In terms of ray propagation, we may interpret these results as implying
that a ray starting from
$\mathbf{x}_0$ at $t_0$ will reach $\mathbf{x}$ at $t$ such that
\beq \label{eq2.23}t-t_0=|\mathbf{x}-\mathbf{x}_0|+\delta t_{GE}+\delta
t_{GM},\eeq
where $\delta t_{GE}$ is the {\it gravitoelectric time delay} given by
\beq \label{eq2.24}\delta
t_{GE}=\int^{\mathbf{x}}_{\mathbf{x}_0}[\mathcal{N}(\mathbf{x}')-1]\, \hat{\mathbf{k}}\cdot
\rmd
\mathbf{x}'.\eeq
For $\Phi <<1$, $\mathcal{N}\sim 1+2\Phi$ and $\delta t_{GE}$ reduces to the integral
of $2\Phi$ along the
straight line from $\mathbf{x}_0$ to $\mathbf{x}$, which is the Shapiro
time delay. Moreover, $\delta
t_{GM}$ is the {\it gravitomagnetic time delay} given by
\beq 
\label{eq2.25} \delta t_{GM}=\int^{\mathbf{x}}_{\mathbf{x}_0}\bfgamma
(\mathbf{x}')\cdot
\rmd\mathbf{x}',
\eeq
which vanishes if $\mathbf{x}-\mathbf{x}_0$ is in the same plane as the
$z$-axis. We note that the
monopole contribution to Eq.~(\ref{eq2.25}), modulo an integer multiple of $8\pi \ell$, vanishes in the equatorial plane.

It has recently been shown that the gravitomagnetic time delay caused by rotational motion could make a significant
contribution to the gravitational lensing delay of extragalactic sources and should therefore be taken into account in the 
interpretation of observational data~\cite{ciufo1,ciufo2,ciufo3}. 
The time delay around a closed loop would involve a gravitoelectric as well
as a gravitomagnetic
component. If the loop is traversed in the opposite sense, the
gravitoelectric time delay for the
stationary field under consideration remains the same but the
gravitomagnetic time delay changes sign. In
this way, one recovers Eq.~(\ref{tpiutmeno}) for the total time delay
$t_+ - t_-$ from a different
standpoint.

Finally, let us consider the influence of the gravitomagnetic monopole on
the polarization of light.
The static background Schwarzschild spacetime has no influence on the polarization of test
electromagnetic radiation due to its spherical symmetry \cite{mas75}. We therefore
assume that $\Phi \ll 1$ and consider
linear perturbations of Minkowski spacetime  such that
$g_{\mu\nu}=\eta_{\mu\nu}+h_{\mu\nu}$,
\beq
h_{00}=2\Phi,\quad h_{ij}=2\Phi \delta_{ij}, \quad h^{0i}=2A^i_{\rm (g)}.
\eeq
Far from the source, $\bfnabla \cdot \mathbf{A}_{\rm(g)}=0$ and 
$\Box
\mathbf{A}_{\rm(g)}=0$, so that the
trace-reversed perturbations
\beq
\bar h_{\mu\nu}= h_{\mu\nu}-\frac12 h \eta_{\mu\nu}
\eeq
satisfy $\bar h{}^{\mu\nu}{}_{,\nu}=0$ and $\Box \bar h^{\mu\nu}=0$. It
then follows from the analysis in
\cite{mas75,kopmas}
that in a gravitomagnetic field $\mathbf{B}_{\rm(g)}$, the plane of linear
polarization rotates at a rate given by \cite{mas75,kopmas}
\beq
   \frac{\rmd \zeta}{\rmd t} = \mathbf{B}_{\rm(g)} \cdot \mathbf{\hat k}\ ,
\eeq
where $\zeta$ is the rotation angle and $\mathbf{\hat k}$ is the unit
propagation vector of the
electromagnetic radiation. This Skrotskii effect \cite{Skr} is the
gravitational analog of the Faraday
effect in electrodynamics. Thus for 
electromagnetic radiation propagating from $\bfrho_1$ to $\bfrho_2$, the plane of
polarization rotates by the angle $\zeta = \Psi(\bfrho_1) - \Psi(\bfrho_2)$ in
this linear approximation. In particular, for
a ray propagating from a radius $\rho_0\gg
M$ out to infinity, the plane of the polarization rotates by an angle
\beq\label{eq2.29} 
\zeta_0 =\Psi_0 
=\frac{\mathbf{J}\cdot \bfrho_0}{\rho_0^3} +\frac{\mu}{\rho_0}\ .
\eeq
This result is consistent with the spherical symmetry of the monopole.
Moreover, it follows from Eq.~(\ref{eq2.29}) that the net angle of rotation of the 
plane of polarization is zero in the scattering case of radiation propagating
from $-\infty$ to $\infty$.

To go beyond the JWKB limit described above, one must consider the gravitomagnetic phenomenon
of helicity-rotation coupling as well; see, for instance, \cite{mas75,car92} for the case of the Kerr spacetime.

In connection with the spin-rotation coupling, let us consider the motion of a classical
spinning test particle in the linearized KTN spacetime given by Eq.~(\ref{eq:TNquasi}). 
In the absence of the gravitational source, we may assume that the free test particle
is at rest and carries a constant spin vector with its direction fixed in the inertial frame.
In the presence of the gravitational source, however, the spin of the particle couples to the spacetime
curvature resulting in the Mathisson-Papapetrou force
\beq\label{eq:FuS}
  F_\alpha = - \frac12 R_{\alpha\beta\gamma\delta} \, u^\beta S^{\gamma\delta}\ ,
\eeq
where $u^\alpha$ is the 4-velocity and $S^{\alpha\beta}$ is the spin tensor of the particle.
For a \lq\lq point\rq\rq $\,$ particle with spin, we need to impose the Pirani supplementary condition
$S^{\alpha\beta} u_\beta=0$ for the sake of completeness. The spin vector of the particle is
defined by
\beq
  S_\alpha = - \frac12 (-g)^{1/2}\epsilon_{\alpha\beta\gamma\delta}\, u^\beta S^{\gamma\delta}
\ .
\eeq

In the linear approximation with $u^\alpha \approx (1,0,0,0)$, $S^{0i} \approx0$, and
$S^{ij} \approx \epsilon^{ijk} S_k$, we find that in Eq.~(\ref{eq:FuS})
the Mathisson-Papapetrou force is $F_0 = 0$ and 
$F_i = -(\mathbf{B}_{\rm(g)})_{j,i} S^j$.
This force may be expressed as the negative gradient of a gravitational spin potential energy
$\mathcal{H} = \mathbf{S}\cdot \mathbf{B}_{\rm(g)}$.
It turns out that this classical result corresponds in the JWKB approximation
to wave-mechanical (quantum) results. Using Eq.~(\ref{eq:Bg}), we see that the energy $\mathcal{H}$
consists of two terms: the familiar gravitational spin-rotation coupling term ($\sim 10^{-28}$eV
for a spin-$\frac12$ particle in an Earth-based laboratory)
and a new term $\mu \mathbf{S}\cdot \bfrho /\rho^3$ that is due to the coupling of spin with
the gravitomagnetic monopole moment of the source. 

\section{Circular holonomy}

Turning now to the parallel transport around closed $\phi$-loops, a key
result of~\cite{MMM} for the
Kerr spacetime involves the relation between the gravitomagnetic temporal
structure around a rotating mass
and circular equatorial holonomy, namely, the holonomy of the time component
$X^t =dt(X)$ of an arbitrary
vector field $X$
does not in general vanish for a spacelike circular
equatorial orbit around a rotating
mass. This temporal holonomy involves a Lorentz boost
and the corresponding time dilation indicates the existence of the
gravitomagnetic clock effect.
The clock effect and the temporal holonomy vanish in the Schwarzschild
spacetime. In the case of the
Taub-NUT gravitational field, the corresponding equation of parallel
transport around a spacelike
circular equatorial orbit implies that
\beq
  \frac{\rmd X^t}{\rmd \phi} + \mu X^\theta =0 \ ,
\eeq
so the temporal holonomy is in general nonzero for $\ell\neq0$. This
indicates that the
gravitomagnetic monopole moment of the source produces a separate temporal
structure around the Taub-NUT
source as indicated, for instance, by the Skrotskii effect discussed above.

The gravitomagnetic clock
effect is essentially absent in this case due to the spherical symmetry of
the monopole source, which is a
gravitational dyon consisting of a gravitoelectric monopole (represented by
the mass $M$) and a
gravitomagnetic monopole (represented by the NUT parameter $\ell$).
Moreover, timelike circular
equatorial geodesic orbits do not exist in the Taub-NUT spacetime. That is, $\ell$
breaks the degeneracy of the
Schwarzschild circular geodesic orbits by lifting them off the equatorial plane in opposite directions. This could be intuitively understood by considering the monopole contribution to the gravitomagnetic Lorentz force per unit mass of the test particle, i.e., $2\mu \bfrho \times \mathbf{v}/\rho^3$~\cite{zee}. To first order in
$\ell$, however, the orbital frequencies are the same as the Keplerian frequencies in the
Schwarzschild spacetime, 
$\rmd\phi/\rmd t = \pm \omega_{\rm(K)}$ with $\omega_{\rm(K)} =
(M/r^3)^{1/2}$. This is
treated in more detail in the following section.

\section{Geodesics in the KNTN spacetime}

The motion of a free uncharged test particle in the Kerr-Newman-Taub-NUT spacetime is given by
\begin{eqnarray}
\fl
\dot r&=& \pm \frac{1}{\Sigma}\sqrt{R}\,\,,\qquad R
=-\Delta(m^2r^2+{\cal
K})+P^2\,\,,\quad P=E(r^2+a^2+\ell^2)-aL_z \ , 
\label{RDOT}\\
\fl 
\dot \theta&=& \pm \frac{1}{\Sigma}\sqrt{\Theta}\,\,,\qquad \Theta
= {\cal K}-m^2(\ell+a\cos\theta)^2-\frac{1}{\sin^2\theta}(L_{z}-E\chi)^2 \ ,
\label{THETADOT}\\
\fl 
\dot \phi&=&\frac{1}{\Sigma}\left [ \frac{1}{\sin^2\theta} \left( L_z-E\chi
\right)+a\,\frac{P}{\Delta}
\right ]\ ,\label{PHIDOT}
\\
\fl
\dot t&=&\frac{1}{\Sigma}\left [
\left(r^2+a^2+\ell^2\right)\frac{P}{\Delta}
  +\frac{\chi}{\sin^2\theta}\left(L_z-E\chi\right)\right ],
\label{TDOT}
\end{eqnarray}
where $m,E,L_z$, and ${\cal K}$ are the four constants of motion for the particle
and represent respectively its rest mass, 
energy, angular momentum about the $z$-axis, and Carter's constant~\cite{Carter}.
Here the dot represents differentiation with respect to the affine parameter $\lambda$  
whose differential is related to the proper time by
$\rmd \tau = m \, \rmd \lambda$ when $m\neq 0$. 
For the timelike orbit of a test particle with nonzero rest mass
$m$, the principle of equivalence implies that the orbit is uniquely characterized
by the reduced constants $\tilde E = E/m$, $\tilde L_z = L_z/m$, and 
$\tilde \mathcal{K} = \mathcal{K} / m^2$. A detailed examination of the solutions of these
equations is beyond the scope of this article; therefore, we shall only consider certain
special orbits here.

Let us first assume that $a=0$. In this case, we choose $L_z\neq0$ and
$\mathcal{K} = (m^2-4 E^2) \ell^2 + L_z^2$; one can show from Eq.~(5.2) that $\theta$ is then fixed
and the orbit is confined to a cone with the opening angle $\theta$ given by $\cos\theta = -2 \ell E/L_z$.  It follows that in this case the equations 
of motion on the cone depend on $\ell$ only via $\ell^2$. As $\ell\to0$, such an orbit reduces to an 
equatorial plane orbit; in particular, for constant $r$ we recover circular equatorial orbits. 
The reflection of the orbit on the cone about the equatorial plane 
can be achieved by simply
changing the sign of $L_z$ since $\rmd \phi / \rmd \tau = L_z/(r^2+\ell^2)$; that is, the sense of
motion is reversed below the equatorial plane. As the gravitomagnetic monopole moment vanishes, the
orbits on the cones with $\cos\theta = \pm |2\ell E/L_z|$ degenerate to a single orbit on the equatorial
plane.

The absence of timelike circular equatorial geodesics extends to the KTN spacetime. More generally, we can look for geodesic orbits that have constant $r$ and $\theta$ coordinates. These would be special circular geodesic orbits around the rotation axis that for $\theta=\pi/2$ coincide with equatorial circular geodesics. A detailed analytic investigation of the geodesic equation shows that for $a\not= 0$ and $\ell \not= 0$ no such (timelike, null, or spacelike) geodesics exist in the KTN spacetime.

Let us next consider timelike spherical geodesics in the KTN spacetime. While such stable orbits
exist all the way down to the horizon (for the corotating orbits with $a = M$) in the Kerr case~\cite{Wil,oneill},
they are bounded away from the horizon of the Kerr-Taub-NUT spacetime, as can be seen by generalizing
Fig.~3 of Wilkins~\cite{Wil} and his related calculations.  To first order in $\ell$, however, the radius of a 
spherical orbit coincides with that of a Wilkins orbit with the same orbital parameters 
$(\tilde E, \tilde L_z, \tilde \mathcal{K})$. In this case the orbit follows a spiraling spherical path that stays 
within latitudes given to first order in $\ell$ by $\theta_{\rm min} = \theta_0 +\theta_1$ and
$\theta_{\rm max} = \pi-\theta_0 +\theta_1$.  Here $\theta_0\in(0,\pi/2)$ is the unperturbed
latitude satisfying
\beq
  (\tilde L_z -a \tilde E \sin^2\theta_0)^2 
 = (\tilde\mathcal{K}- a^2 \cos^2\theta_0) \sin^2\theta_0 \ .
\eeq

If  $\theta_0$ is a solution of this equation, then so is $\pi -\theta_0$; therefore, the 
Wilkins orbits are symmetric in  latitude about the equatorial plane. The first order perturbation
in latitude $\theta_1$ is obtained from $\Theta=0$ and can be expressed as
\beq
  \theta_1 = \left(\frac{\ell}{\sin\theta_0}\right)
          \frac{2\tilde E \tilde L_z + a (1-2 \tilde E {}^2)\sin^2\theta_0}
               {\tilde\mathcal{K} +2a\tilde E \tilde L_z -a^2+ 2 a^2(1-\tilde E{}^2)\sin^2\theta_0}
\ .
\eeq
Thus to linear order in the strength of the gravitomagnetic monopole moment, the spherical orbits are
shifted up or down and are no longer symmetric in latitude about the equatorial plane.

\section{Discussion}

Following up on our previous work \cite{bjm}, we have considered some of the main effects
associated with the existence of a gravitomagnetic monopole moment. 
The time coordinate $t$ is periodic in the KNTN spacetime with a period that is simply proportional
to the gravitomagnetic monopole moment of the source. The gravitational time delay and the rotation of the plane of polarization of electromagnetic waves studied in this paper further elucidate the special temporal structure in this spacetime.  We have also briefly studied the influence of the gravitomagnetic monopole moment on the motion of a spinning particle, and on the spherical orbits in KTN spacetime generalizing some of the previous results of Wilkins~\cite{Wil}.  
There is no observational evidence at present for the existence of a gravitomagnetic monopole~\cite{ZimSha,LynNou}.
Our work could
in principle be combined with the analysis of astronomical data in order to set upper limits on the possible
existence of gravitomagnetic monopoles.

\end{document}